\documentclass[showpacs,preprint,amssymb]{revtex4}
\usepackage{graphicx}
\usepackage{dcolumn}
\usepackage{bm}
\def\be{\begin{equation}} 
\def\ee{\end{equation}}
\def\bea{\begin{eqnarray}} 
\def\eea{\end{eqnarray}}
\def\line{\hbox to \hsize}    
\def\frac #1#2{{#1\over #2}}

\def\tr{{\rm  tr\,}}

\def\psid{\psi^{\dagger}}

\def\bz{{\bar z}}
\def\bepsilon{{\bar \epsilon}}
\def\bzeta{{\bar\zeta}}

\def\sgn{{\rm sgn\,}}

\def\e{{\bf e}}

\def\vev #1{{\langle #1\rangle}}
\def\1{\mbox{\bf 1}}

\begin{document}

\title{Gravitational Anomalies and Thermal Hall effect in Topological Insulators}

\author{ MICHAEL STONE}

\affiliation{University of Illinois, Department of Physics\\ 1110 W. Green St.\\
Urbana, IL 61801 USA\\E-mail: m-stone5@illinois.edu}   

\begin{abstract}  

It has been  suggested that a temperature gradient will induce a Leduc-Righi, or thermal Hall, current   in the Majorana quasiparticles   localized on  the surface of class DIII topological insulators, and that the    magnitude of this current  can be related {\it via\/} an Einstein argument to a Hall-like energy flux induced by gravity.  We  critically examine this idea, and argue  that the gravitational  Hall effect is more complicated than its familiar analogue.  A conventional Hall current is generated by a {\it uniform\/}  electric field,  but  computing the  flux from the gravitational Chern-Simons functional shows that  gravitational field {\it gradients\/} ---  {\it i.e.} tidal forces --- are needed  to induce a  energy-momentum  flow.  We   relate the surface energy-momentum  flux to a domain-wall gravitational anomaly {\it via\/} the   Callan-Harvey  inflow mechanism. We  stress that  the gauge invariance of the combined bulk-plus-boundary theory   ensures that the current in the domain wall     always  experiences  a  ``covariant''   rather than   ``consistent'' anomaly. We  use this  observation  to  confirm  that  the tidally induced energy-momentum current  exactly accounts for the covariant gravitational anomaly in $(1+1)$ dimensional domain-wall fermions. The same anomaly arises whether we write the Chern-Simons functional in terms of the Christoffel symbol or in terms of the  the spin connection. 

\end{abstract}


\maketitle

\section{Introduction}

One of the key properties  of topological insulators is the intimate connection between the non-trivial bundle structure of the bulk electronic states and the presence  of protected gapless surface modes. The most intuitive way of understanding this  connection is that the twisted  bundle  gives rise to bulk quantum-Hall-like conductivities,   and the  gapless surface modes   need  to be present   to soak up  the corresponding  conserved currents where     they run into the surface  of the sample \cite{boyanovsky-dagotto-fradkin,stone_hall_edge}.  In this way the bulk-surface connection is seen to be  a manifestation  of the Callan-Harvey ``anomaly inflow''  mechanism \cite{callan-harvey}.   Most of the Altland-Zirnbauer classes \cite{zirnbauer1,zirnbauer2} of topological insulators  possess conserved ${\rm U}(1)$  charge or ${\rm SU}(2)$ spin currents,  and the necessity of their protected surface modes can be understood {\it via} ordinary gauge-field anomalies.  An important exception is the class  DIII, which includes superconductors with spin-orbit interactions, and superfluid $^3$He-B. Here the only conserved quantities  are energy and (in the translation invariant superfluid) momentum. 
An anomaly-inflow understanding of the electrically neutral (Majorana) surface  modes in the DIII  systems therefore requires a failure of  some edge-mode energy-momentum conservation law --- in other words a {\it gravitational anomaly\/} \cite{alvarez-gaume-witten}.

Gravitational anomalies   originate in the $\hat A$-genus contribution to the Dirac index theorem that is non-zero only in  $4k$ space-time dimensions. They  descend {\it via\/} a parity-violating gravitational Chern-Simons term in $4k-1$ dimensions  to  an energy-momentum  inflow anomaly in $4k-2$ space-time dimensions.  For physically realizable topological insulators we are restricted to the $k=1$, and therefore to a gravitational Chern-Simons term in a $(2+1)$-dimensional surface, and a gravitational anomaly in a $(1+1)$-dimensional edge.  

Following \cite{vurio,goni,pisarski,volovik,read-green} we expect that (after the application of a small symmetry-breaking field that opens a gap)  the  $(2+1)$-dimensional Majorana fermion surface modes of  the DIII systems will  include  a  gravitational Chern-Simons term in their low-energy effective action.
It is  argued  in  
\cite{ryu-moore-ludwig, wang-liang-zhang, nomura-ryu-furusaki-nagaosa}
that this  term can, in principle,  be observed  through a   thermal  Hall (or Leduc-Righi) effect.  A key step in the  reasoning  in \cite{ryu-moore-ludwig,wang-liang-zhang,nomura-ryu-furusaki-nagaosa}  requires  that, in  analogy with the conventional Hall effect,   a  uniform gravitational field   induces a   surface energy-momentum current.  The Leduc-Righi, coefficient is then obtained  by means of  an  Einstein argument.   The idea is that  thermal equilibrium in  the presence of a gravitational field requires the local temperature to vary so as to compensate  for the  gravitational red shift experienced by radiation   as it moves in the potential. The  energy flux  induced by a  thermal gradient  is then balanced by an equal and opposite energy   flux due to the gravitational potential gradient. The thermal Hall conductance  can thus  be found from  that of the gravitational Hall  conductance.   

The purpose of this paper   is to argue that,  although the arguments in \cite{ryu-moore-ludwig, wang-liang-zhang,nomura-ryu-furusaki-nagaosa}
are  very appealing, the gravitational  ``Hall effect''  is a little more complicated than its electromagnetic  analogue.  While  a temperature gradient across a finite $(2+1)$-dimensional surface does indeed induce a thermal Hall current whose magnitude  is related to the gravitational anomaly \cite{read-green,cappelli-huerta-zemba}, the surface-state  energy gap exponentially  suppress any surface  thermal  current. The heat  must therefore be  carried entirely by the $(1+1)$-dimensional edge modes.  In this respect the thermal current differs from the charge Hall effect, which  can flow either at the $(1+1)$-dimensional edge, or,  in the presence of a uniform electric field, within  the $(2+1)$-dimensional electron gas.     Furthermore,   the  gravitational Chern-Simons term yields  an   energy-momentum flux that is   proportional to to gradients of the Ricci tensor.   Consequently a    {uniform\/}  bulk  gravitational field  cannot  create an energy-momentum flux {within\/}  the $(2+1)$-dimensional surface.  A  surface energy flux requires an {\it inhomogeneous\/}  field --- {\it i.e.\/}\   {tidal\/} forces.  Nonetheless,  the  tide-induced energy-momentum flow does   retain  the bulk-boundary connection because it demands  an anomalous   $(1+1)$-dimensional  gapless mode to absorb the flux  as it   runs into  an edge  or   domain wall.     

In section   \ref{SEC:Hall} we will describe the thermal Hall effect and show how it can maintain  an equilibrium balance  between a temperature gradient and a gravitational potential gradient even in the absence of a bulk energy flow. In  section \ref{SEC:inflow} we review the Callan-Harvey  anomaly-inflow picture, and stress that this  mechanism always leads to the {\it covariant\/} form of the associated anomaly. In section \ref{SEC:2Dgrav} we explain the origin of the gravitational anomaly in $(1+1)$ dimensional chiral theories. In section \ref{SEC:gravCS} we  compute the energy momentum flows arising from a $(2+1)$-dimensional gravitational Chern-Simons functional and show that it exactly accounts for the anomaly obtained in section \ref{SEC:2Dgrav}.  We also show that the same anomaly is obtained from the Chern-Simons functional whther it is  written in terms of the Christoffel symbols $\Gamma$ or the spin connection $\omega$.  Finally section \ref{SEC:conclusions} provides a brief summary of our results.

\section{Thermal Hall currents}
\label{SEC:Hall}

The edge of a (2+1)-dimensional quantum Hall system hosts  gapless chiral fermions  \cite{wen-edge}, and   both the edge of  a $p_x+ip_y$ superconductor and domain walls on the surface   of a suitably engineered topological insulator host chiral {\it Majorana\/}   fermions  \cite{read-green,fu-kane}.   Suppose for a moment that  these modes can be modelled  as a set of $n$ independent conformal fields  having (positive or negative)  propagation velocities $v_i$, $i=1,\ldots,n$ and central charges $c_i$. Then, at temperature $T$, each independent chiral edge-mode contains  an energy density \cite{affleck}
\be
\varepsilon_i = c_i \frac{\pi}{12|v_i| } \frac{k^2_B}{\hbar} T^2,
\ee
where $k_B$ is the Boltzmann constant.
Thermal energy is therefore being transported along the edge at a rate \cite{kane-fisher}
\bea
J_T&=&\sum_{i=1}^n v_i \varepsilon_i\nonumber\\
&=&\frac{\pi}{12} \sum_{i=1}^n \sgn(v_i) c_i  \frac{k^2_B}{\hbar} T^2\nonumber\\
&=& \frac{\pi}{12}  (c-\bar c) \frac{k^2_B}{\hbar} T^2.
\label{EQ:edge_energy_flux}
\eea
Here $c$ and $\bar c$ are the total conformal charges of the right and left-moving modes respectively. Although motivated by the model of  independent modes, this formula continues to hold for  more complicated  conformal edge-mode theories \cite{cappelli-huerta-zemba}.   If we construct a parallel-sided Hall bar and maintain a small temperature gradient $\Delta T$ across it, then the difference between  the 
contra-propagating energy fluxes (\ref{EQ:edge_energy_flux})  on  the two edges gives rise to a  net thermal current 
\be
J_{\rm L-R}=C_{\rm L-R}\Delta T
\ee
that flows along the bar and perpendicular to the temperature gradient.  Here
\be
C_{\rm L-R} = (c-\bar c)\frac{\pi}{6} \frac{k^2_B}{\hbar} T, 
\ee
is  the   Leduc-Righi  coefficient.

It is remarkable that the 
non-universal   edge-mode velocities have cancelled, leaving in $C_{\rm L-R}/T$ only fundamental constants  and the numbers $c$, $\bar c$ that are characteristic of the quantum Hall phase. It is therefore reasonable to suppose that  $C_{\rm L-R}/T $  may be extracted from  topological data, as is the quantum Hall coefficient.   It is, however,  difficult to provide a  direct  derivation of thermal conductivities from linear response theory. There is no term  that can be added to the Hamiltonian to describe the temperature.   An ingenious trick was introduced by Luttinger \cite{luttinger}, who instead coupled the system to {\it gravity\/} and proceeded indirectly by adopting the method used by Einstein to relate diffusion coefficients to viscosity \cite{einstein}.
Luttinger's  idea is that the de-equilibrating effect of a small  temperature gradient will  be precisely compensated  for by a gravitational potential $\Phi$ 
provided that
\be
\frac{1}{T}\frac{\partial T}{\partial x}= -\frac{1}{c_{\rm light}^2}\frac{\partial \Phi}{\partial x}.
\label{EQ:weak_gradient}
\ee  
Consequently,  assuming that all currents vanish in equilibrium, and that the effects of the two driving forces are additive,  a linear-response derivation  of  the current induced by gravity allows one to find the current induced by the thermal gradient.

In the present case  imagine a rectangular Hall bar whose upper,  right-propagating,  edge at co-ordinate $y=y_1$ is held at temperature $T_1$  and whose left-propagating lower edge at $y=y_2$ is held at temperature $T_2>T_1$.   If the  bar lies  in a gravitational field  such that the gravitational frequency    shift obeys
\be
\frac{\omega(y_1)}{\omega(y_2)}\equiv  \sqrt{\frac{g_{00}(y_2)}{g_{00}(y_1)}}= \frac{T_1}{T_2},
\ee   
then, as the thermal excitations from  the hotter lower edge   rise on the left hand vertical side to the upper edge they  will red-shift  to the lower temperature. 
 Similarly, as the excitations from  the cooler upper edge descend {\it via\/} the right hand vertical side to the lower  edge they will  blue-shift to match the hotter  temperature. The system is in equilibrium therefore. Since for weak gravitational fields we have
\be
\sqrt{g_{00}(y)}\approx 1 +\frac{\Phi(y)}{c_{\rm light}^2},
\ee
this situation satisfies (\ref{EQ:weak_gradient}).   Observe, however, that in our Hall bar, the currents are {\it not\/} zero in equilibrium. Therefore a knowledge of the thermal Hall current  does not allow one to deduce  the gravitational Hall current, nor  {\it vice versa\/}.   

The steady-state  equilibrium of the Hall bar requires  no thermal or other form of energy to be flowing  within the gapped surface states. Indeed  an extra energy-momentum flux  into the gapless states on the vertical sides would mess things up. This already suggests that a uniform gravitational field should  not induce a surface  energy flow. 

This suggestion is  perhaps not surprising. A mathematical  analogy between the conventional Hall effect and gravitation would naturally identify the field strength $F$ with the Riemann curvature $R$.  A  uniform field gravitational field does not, however,   require  space-time  curvature. The   Rindler metric
\bea
d\tau^2&=& \left (1+\frac {(r-r_0)g}{c_{\rm light}^2}\right)^2 dt^2-\frac1{c_{\rm light}^2}dr^2,\quad g= c_{\rm light}^2/r_0 
\eea
of  a uniformly accelerated observer provides a gravitational potential
 $\Phi(r)=(r-r_0)g,$ but is merely  a re-parametrization
\bea
c_{\rm light}T&=& r\sinh \left(\frac{c_{\rm light}t}{r_0}\right),\nonumber\\
X&=& r\cosh\left(\frac{c_{\rm light}t}{r_0}\right),\nonumber
\eea
of a part of Minkowski space with flat metric
\be
 d\tau^2= dT^2-\frac{1}{c_{\rm light}^2}dX^2.
\ee
It might, therefore, be more physical  to identify the thermal-Hall analogue of the  electric  field with the Christoffel symbols $\Gamma$ which  describe the frame-dependent inertial forces  that we perceive as gravity.  If this new analogy is to work,  the energy-momentum influx into the edge modes would  have to be given by the non-covariant ``consistent''  gravitational  anomaly,  which contains $\Gamma$'s, rather that the ``covariant'' anomaly which  contains only $R$ \cite{bardeen-zumino}.
In the following sections, however,  we will argue    that  anomaly inflow always give rise to the {\it covariant\/} anomaly, and not to the consistent anomaly. Moreover, we will   see that {\it gradients\/} of  curvature  are needed to  produce an energy  flow into edge states.  

To simplify the argument, we will follow the authors of
\cite{ryu-moore-ludwig} and argue that   since we are interested in topological effects, we can choose non-universal quantities such as propagation velocities as we like. We will therefore from now on make all modes propagate at the speed of light, work with fully relativistic systems, and use  natural units, in which $\hbar=c_{\rm light}=1$.

\section{The Callan-Harvey mechanism and covariant versus consistent anomalies} 
\label{SEC:inflow}

Let us recall how the conservation (or non-conservation) of a gauge  current is related to the gauge  invariance (or the lack of it)  of an   action functional. Suppose, for example, that   $S[A]$  is a functional of  an  $\mathfrak{su}(N)$   Lie-algebra-valued gauge field  $A_\mu = \lambda_a A^a_\mu$, where the matrices $\lambda_a$ are the generators of $\mathfrak{su}(N)$.
We define the matrix-valued  gauge current  $J^\mu(x)= \lambda_a J^{\mu,a} $  by setting 
\be
\delta S[A]= \int d^dx \, \tr\{ J^\mu \delta A_\mu\}.
\ee
Under a gauge transformation the field changes as $A_\mu \to A_\mu^g =g^{-1}A_\mu g +g^{-1}\partial_\mu g$, where $g\in {\rm SU}(N)$.  For  an infinitesimal transformation  $g= 1-\epsilon$ 
the transformation  becomes  $A_\mu \to A_\mu+\delta_\epsilon A_\mu$, where $\delta_\epsilon A_\mu=- ([A_\mu, \epsilon]+\partial_\mu \epsilon)\equiv - \nabla_\mu \epsilon $. The corresponding change in $S[A]$ is
\bea
\delta_\epsilon S &=& \int d^dx\,  \tr\{ J^\mu ( [\epsilon, A_\mu]-\partial_\mu \epsilon)\}\nonumber\\
&=& \int d^dx\, \tr\{\epsilon (\partial_\mu J^\mu +[A_\mu, J^\mu])\}.
\eea
The covariant divergence
\be
\nabla_\mu J^\mu \equiv \partial_\mu J^\mu+[A_\mu, J^\mu]  
\ee
is therefore  zero if and only if  $S[A]$ is  gauge invariant.

We are interested in {\it effective actions\/}  $S[A]$  that arise as a result    integrating out  a collection of  Fermi fields  $\psi$, $\psid$ in the presence of a classical background gauge field $A_\mu$: 
\be
\exp\{-S[A]\}= \int d[\psi]d[\psid] \exp\left\{ - S[\psi,\psid,A]\right\}.
\ee
The calculated currents  are then the  expectation value
$J^\mu =\vev{\hat J_\mu}$ of a quantum operator. 
 The original $S[\psi,\psid,A]$ action will  be invariant under $A_\mu\to A_\mu ^g$, $\psi\to g^{-1}\psi$, $\psid \to \psid g$, but the invariance may  lost during the functional integration. In this case we will have
\be
\nabla_\mu J^\mu=G(A),
\ee
where  the {anomaly\/}  $G(A)$  is a local polynomial in the $A_\mu$ and their derivatives.  A gauge anomaly provides an obstruction to a   subsequent quantization of  the $A_\mu$ fields, but 
when  the $A_\mu$ are  simply classical probes it provides a   useful source of non-perturbative information.

The Callan-Harvey effect links the non-conservation of gauge and other currents  to an inflow of charge from some higher dimensional space in which the anomalous theory is embedded as  modes localized on a  domain wall or string defect.  In the cases we are interested in, the inflow is derived from a Chern-Simons term in one-higher space dimension.   

As usual we will   think of $A$ as a Lie-algebra-valued one-form $A=\lambda_a A^a_\mu dx^\mu$, and define the field strength as the Lie-algebra-valued two-form
\be
F=dA+A^2= \frac 12 F_{\mu\nu}dx^\mu dx^\nu.
\ee
The  Chern-Simons form $\omega_{2n-1}(A)$  is then defined as   
\be
\omega_{2n-1}(A)= n\int_0^1 \tr\{ AF^{n-1}_t\},
\ee
where $F_t =tF+t(t-1) A^2$.
It is constructed so  that $d\omega_{2n-1}= \tr\{F^n\}$.  
For example,
\bea
\omega_3(A)&=& \tr\{AdA+\frac 23 A^3\},\nonumber\\
&=& \tr\{AF-\frac 13 A^3\},
\eea
and
\bea
\omega_5(A)
&=& \tr\{ A(dA)^2 +\frac 32 A^3 dA +\frac 35 A^5\}\nonumber\\
&=& \tr\{ AF^2-\frac 12 FA^3 +\frac 1{10} A^5\}.
\eea
The $F$-free  last term $\propto A^{2n-1}$ in the second forms of $\omega_{2n-1}$ has coefficient
\be
c_n=(-1)^{n-1}\frac{n! (n-1)!}{(2n-1)!}.
\ee
It is this last term that governs the change in integrals of $\omega_{2n-1}$ under large gauge transformations. 
If $A$  undergoes a finite gauge transformation
\be
A\to A^g =g^{-1}A g+g^{-1}dg,
\ee
then 
\be
\omega_{2n-1}(A^g)= \omega_{2n-1}(A)+ c_n \tr\{(g^{-1}dg)^{2n-1} \}+d \alpha_{2n-2}(A,g),
\ee
where, for example \cite{manes}
\be
\alpha_2= - \tr\{dg g^{-1}A\}
\ee
and
\be
\alpha_4(A,g) = -\frac 12 \tr\{(dgg^{-1})(AdA+dAA+A^3) -\frac 12 (dgg^{-1}) A(dgg^{-1}) A-(dgg^{-1})^3 A\}.
\ee

The Chern-Simons  functional $C[A]$ is defined by setting  
 \be
C[A]=2 \pi   \left(\frac{i}{2\pi}\right)^{n}\frac 1 {n!}\int _M \omega_{2n-1}(A),
\ee
where $M$ is some $2n-1$ dimensional manifold.  The coefficient in front of the integral has been chosen so that  $\exp\{i C[A]\}$  is single-valued when  $M$ is the  $(2n-1)$-sphere.  In this case 
\be
C[A^g]-C[A] =2\pi  \left(\frac{i}{2\pi}\right)^{n}\frac{(n-1)!}{(2n-1)!} \int_{S^{2n-1}} \tr\{(g^{-1}dg)^{2n-1}\},
\label{EQ:winding_number}
\ee
 and it is shown in \cite{bott-seeley} that the  right-hand side of  (\ref{EQ:winding_number})  is  $2\pi $ times an integer  whenever   $g \in {\rm GL}(n,{\mathbb C})$ or any of its compact subgroups such as ${\rm SU}(N)$.    This means that when a Chern-Simons functional appears in a functional  integral
 \be
 Z= \int d[A] \exp\{ikC[A]+\cdots\}
 \ee
 then gauge invariance demands that $k$  be an integer.  This  constraint on $k$ need not hold when $C[A]$ appears in an effective action. Indeed $k$ is $1/2$  when we integrate   out a massive  Dirac fermion  in odd dimensional  space time.

Given a $(2n-1)$-manifold $M$ possessing  a $2n-2$-dimensional boundary $\partial M$,  we can use $C[A]$ to construct an action
$
S[A,g]  =C[A^g]
$
that is obviously invariant under $A\to A^h$, $g\to h^{-1}g$. In this action, the  gauge non-invariance of the bulk Chern-Simons term  $C[A]$ is compensated by the complementary gauge non-invariance of the {Wess-Zumino action\/} 
\bea
W[A,g]&\stackrel{\rm def}{=}&  C[A^g]- C[A]\nonumber\\
&=& 2\pi   \left(\frac{i}{2\pi}\right)^{n} \frac {1}{n!} \left\{\int_{\partial M} \alpha_{2n-2}(A,g)+ c_n \int_M \tr\{(g^{-1}dg)^{2n-1}\}\right\}.
\eea
Although $W[A,g]$ requires $g$ to be defined  on the $2n-1$ dimensional manifold $M$, the  identity 
\be
\delta \tr  \{(g^{-1}dg)^{2n-1}\}=  (2n-1)  d \tr\{(g^{-1} \delta g)(g^{-1} d g)^{2n-2}\}
\ee
ensures that 
variation of   $W[A,g]$  depends only on the values that $\delta A$ and $\delta g$ take on the boundary $\partial M$.  It can therefore serve as an anomaly-capturing  non-local  effective action for a $2n-2$ dimensional theory  \cite{witten}.  The  meaning of the gauge-group element $g$ depends on the context.    In a two dimensional  boundary $g(x,t)$ could  be the dynamical chiral boson equivalent to  a chiral fermion.  In two or higher dimensions it  might  parametrize a  Higgs field that gives a left-handed chiral fermion a mass by coupling it to a right handed chiral fermion that does not itself couple to $A$.

The gauge anomaly in a  Wess-Zumino action for a four dimensional theory may be read off from 
\bea
\int_{\partial M} d^4x \, \tr \{\epsilon \nabla_\mu J^\mu_{\rm WZ}\} &=&\delta_\epsilon W[A,g]\nonumber\\
&=& -\delta_\epsilon C[A]\nonumber\\
&=&-\frac 1{24\pi^2}\int_{\partial M} \tr\{d \epsilon(AdA+dAA +A^3)\} \nonumber\\
&=& \frac 1{24\pi^2}\int_{\partial M}  \tr\{ \epsilon \partial_\mu (A_\nu \partial_\sigma A_\tau + \partial_\nu A_\sigma A_\tau+A_\nu A_\sigma A_\tau)\}
\varepsilon^{\mu\nu\sigma\tau} d^4x.
\eea
So
\be
 \tr \{\epsilon \nabla_\mu J^\mu_{\rm WZ}\} =\frac 1{24\pi^2} \tr\{ \epsilon \partial_\mu (A_\nu \partial_\sigma A_\tau + \partial_\nu A_\sigma A_\tau+A_\nu A_\sigma A_\tau)\}
\varepsilon^{\mu\nu\sigma\tau}.
\ee
Because this anomaly is found as the variation of the   functional $W[A,g]$, it satisfies the {Wess-Zumino consistency condition\/}
$$
(\delta_\epsilon\delta_{\epsilon'}- \delta_{\epsilon'}\delta_{\epsilon})W =\delta_{[\epsilon,\epsilon']}W.
$$
It is  therefore known as a ``consistent'' anomaly. The right hand side of the (non) conservation equation is not gauge covariant, however, and so neither is the left. The gauge current itself  is therefore not covariant, and the physical meaning of the (non) conservation equation is unclear.  

In the full  bulk-plus-boundary theory, whose gauge-invariant effective action  is  $C[A^g]$  the   non-zero divergence  of the boundary current is being supplied by the inflow of  gauge current from the higher dimensional bulk. This bulk current {\it is\/} covariant, 
\be
\tr\{ \lambda_a J^\lambda\}= \frac 1{32 \pi^2} \tr\{\lambda_a F_{\mu \nu}F_{\sigma\tau}\} \varepsilon^{\lambda\mu\nu\sigma\tau}.
\ee
It 
comes  from the variation
\bea
\delta \int \omega_5&=&3 \int_M \tr \{\delta A F^2\} + \int_{\partial M}\tr\{\delta A(AdA+dAA+\frac 32 A^3)\}\nonumber\\
&=& 3 \int_M \tr \{\delta A F^2\} + \int_{\partial M}\tr\{\delta A(AF+FA-\frac 12 A^3)\}.
\eea
We usually ignore the boundary term when computing a bulk current, but in the total bulk-plus-boundary theory we must retain it as  it provides a contribution to the  current in the boundary of
\be
\tr\{\lambda_aX^\mu\} \stackrel{\rm def}{= }\frac 1{48 \pi^2} \tr\{\lambda_a(A_\nu F_{\sigma\tau}+ F_{\nu\sigma}A_\tau - A_\nu A_\sigma A_\tau)\}\varepsilon^{\mu\nu\sigma\tau}. 
\ee
This quantity is exactly the extra current (\cite{bardeen-zumino}  equation (2.16)) that has to be added to the consistent current to obtain the {\it covariant\/} anomaly
\be
 \tr \{\lambda_a  \nabla_\mu (J^\mu_{\rm WZ}+X^\mu) \}=  \frac 1{32 \pi^2} \tr\{\lambda_a F_{\mu \nu}F_{\sigma\tau}\} \varepsilon^{5 \mu\nu\sigma\tau}.
\ee
The new current $J_{\rm tot}^\mu=J^\mu_{\rm WZ}+X^\mu$ is now gauge-covariant, and its anomalous divergence entirely accounted for by the Callan-Havey anomaly inflow \cite{naculich,harvey-ruchayskiy}.  

Similarly, in  two dimensions we find that
\be
\nabla_\mu J^\mu_{\rm WZ}
=  \frac{1}{4\pi} \epsilon^{\mu\nu} \partial_\mu A_\nu
\ee
is  the consistent anomaly, and 
\be
X^\mu= \frac{1}{4\pi} \epsilon^{\mu\nu}A_\nu
\ee
is  the Chern-Simons term's contribution  to the boundary current. Then
\bea
\nabla_\mu(J^\mu_{\rm WZ}+X^\mu)&=&   \frac{1}{4\pi} \epsilon^{\mu\nu} \partial_\mu A_\nu+\frac{1}{4\pi} \epsilon^{\mu\nu}(\partial_\mu A_\nu +[A_\mu,A_\nu])\nonumber\\
 &=& \frac{1}{4\pi}\epsilon^{\mu\nu}(\partial_\mu A_\nu-\partial_\nu A_\mu +[A_\mu, A_\nu])\nonumber\\
 &=&  \frac{1}{4\pi}\epsilon^{\mu\nu}F_{\mu\nu},
 \eea
 is the covariant anomaly.

We have seen that   {Bardeen-Zumino\/} polynomial $X^\mu(A)$  that converts the consistent gauge current to the covariant gauge current is precisely the contribution to the boundary   current provided by  the  boundary variation of the bulk Chern-Simons functional.  
   The analogous conversion of a consistent to a  covariant {\it gravitational\/}  anomaly requires an extra integration by parts, and so is more intricate. Indeed some puzzlement was expressed in \cite{callan-harvey} about what happened to the inflowing energy-momentum --- see the discussion  after  equation (30) in  \cite{callan-harvey} --- but it was later understood that the anomaly inflow is always leads to a   covariant current  \cite{naculich, harvey-ruchayskiy}.  

In the above examples, the Chern-Simons term was defined in the bulk and the lower-dimensional  degrees of freedom resided on the boundary. This is, for example, the situation  in the ordinary quantum Hall effect. For $(3+1)$-dimensional topological insulators it is the Chern-Simons functional that is defined on the boundary, and the lower-dimensional theory is defined on a domain wall within the boundary.  In this case the coefficient  of the Chern-Simons functional is multiplied by ${\rm sgn}(m)/2$, where $m$  denotes the   mass gap induced by a  small symmetry breaking field that changes sign at the domain wall.  The resulting domain-wall chiral fermions  then experience half of the the usual   inflow from each side, but there are {\it two\/} sides, and so the resulting edge-theory anomaly is unchanged.

\section{Two-dimensional gravitational anomalies}
\label{SEC:2Dgrav}

In this section we will review  the origin and possible forms of gravitational anomalies.
We start from an effective action  $S[g]$ that depends on the space-time  metric $g_{\mu\nu}$. The associated  Hilbert   
  energy-momentum tensor $T^{\mu\nu}$ is then defined by the variation
\bea
\delta S_{\rm eff}&=&- \frac 12 \int d^dx \sqrt{g} \,T^{\mu\nu}\,\delta g_{\mu\nu}, \\
&=& + \frac 12 \int d^dx \sqrt{g} \,T_{\mu\nu}\,\delta g^{\mu\nu}.
\eea
Under a change of co-ordinates $x^\mu\to x'^\mu =x^\mu+\epsilon^\mu$ we have $g_{\mu\nu}\to g'_{\mu\nu}= g_{\mu\nu}+\delta g_{\mu\nu}$, where 
\bea
\delta g_{\mu\nu}&=& \left({\mathcal L}_\epsilon g\right)_{\mu\nu}\nonumber\\
&=&\nabla_\mu \epsilon_\mu +\nabla_\nu\epsilon_\mu.  
\eea Here  ${\mathcal L}_\epsilon g$ denotes  the Lie derivative of the metric with respect to $\epsilon^\mu$, and $\nabla_\mu$ is the covariant derivative with respect to the torsion-free Levi-Civita connection.
 When the effective action is invariant under this reparametrization we  find (taking into account that  $T^{\mu\nu}=T^{\nu\mu}$) that 
\bea
0 &=&-  \int d^dx \sqrt{g}  \,T^{\mu\nu}\, \nabla_\mu\epsilon_\nu\nonumber\\
&=&  \phantom{-}\int d^dx \sqrt{g}  \epsilon_\nu \nabla_\mu T^{\mu\nu}.
\eea
Thus a gravitational anomaly ---  {\it i.e.} a failure of the covariant conservation law $\nabla_\mu T^{\mu\nu}=0$ --- reflects a failure of reparametrization invariance. While it seems reasonable  that any  physical system should be independent of how we choose  to describe it,  co-ordinate dependence can creep into $S[g]$   when  we tacitly tie a regularization procedure to  the co-ordinate grid  rather than to some intrinsic property such as  the metric.

An equivalent {\it Lorentz anomaly\/} can also occur in theories when  we use a frame field  $e^\mu_a$ rather than the metric to encode the geometry. This anomaly manifests itself as a failure of the energy momentum tensor (now defined in terms of a functional derivative with respect to $e^\mu_a$) to be symmetric.       

We will focus on two-dimensional systems expressed in terms of Euclidean signature isothermal co-ordinates $x,y$, in which 
$
ds^2 =e^\phi (dx^2+dy^2).
$
It is convenient to set $z=x+iy$, $\bz=x-iy$ so that $ds^2=e^\phi dz d\bz$.
The non-zero component of the metric tensor and its inverse are then  $g_{\bz z}=g_{z\bz}= (1/2)e^{\phi} $, and  $g^{\bz z}= g^{z\bz}=2e^{-\phi}$.
In these complex isothermal  co-ordinates the only non-zero entries in the Levi-Civita connection  are 
\bea
\Gamma^z_{zz} &=& \partial_z\phi,\nonumber\\
\Gamma^{\bz}_{\bz \bz}&=&\partial_\bz \phi.
\eea
The  curvature is completely encoded in the Ricci scalar 
\bea
R= {R^{\mu\nu}}_{\mu\nu}={R^{\bz z }}_{\bz z}+ {R^{z\bz}}_{z\bz} = - 4 e^{-\phi} \partial^2_{z\bz}  \phi.
\eea
In our convention $R$ is twice the Gaussian curvature, and hence positive for a sphere. 

The effective action for a left-right symmetric theory with conformal central charge $c$ was obtained  by Polyakov \cite{polyakov} as 
\bea
S_{\rm Polyakov}[g]&=& -\frac c{96\pi}\int d^2x \,(\partial \phi)^2\nonumber\\ 
&=& -\frac c{24\pi} \int  d^2x\, \partial_z\phi \partial_\bz \phi.
\label{EQ:polyakov}
\eea
Here $d^2x$ denotes $dx\wedge dy= d\bz \wedge dz/2i$. To   evaluate   (\ref{EQ:polyakov}) for a given geometry we must select a system of isothermal co-ordinates, and this choice is not unique.  It is therefore not immediately obvious that   $S_{\rm Polyakov}[g]$ is co-ordinate independent.  To verify that it is so, we must examine the conservation of the  energy-momentum tensor. 

Now to  make use of the Hilbert definition of $T^{\mu\nu}$, we must be free to make an arbitrary infinitesimal   variation in the metric.  A general  variation, however, will take us away from the class of isothermal metrics. We therefore make a variation $\delta g_{\mu \nu}$ and follow if with a change of co-ordinates 
\bea
z&\to&z'= z+\epsilon(z,\bz)\nonumber\\
\bz&\to& \bz'=\bz + \bepsilon(z,\bz)
\eea
so as to return to the isothermal gauge. Now 
\bea
\delta(ds^2) &=& [e^\phi(\epsilon\partial_z \phi+\bepsilon\partial_\bz \phi +\partial_z\epsilon+\partial_\bz \bepsilon)+\delta g_{\bz z}+\delta g_{ z\bz})] d\bz dz\nonumber\\
&&\quad\quad  + (\delta g_{zz} +e^\phi \partial_z\bepsilon)dzdz+ (\delta g_{\bz\bz} +e^\phi \partial_\bz\epsilon)d\bz d\bz.
\eea
The required co-ordinate change is obtained by   solving  
\bea
e^\phi \partial_z \bepsilon &=&-\delta g_{zz}\nonumber\\
e^\phi \partial_\bz \epsilon &=&-\delta g_{\bz \bz}.
\label{EQ:finding_epsilon}
 \eea
Let us  assume for the moment that given $\delta g_{zz}$ and $\delta g_{\bz \bz}$ we can always solve these equations for $\epsilon$ and $\bar\epsilon$.  Then,   comparing  with $\delta(ds^2)= e^\phi \delta \phi d\bz dz$ we find that the metric variation leads to  
 \be
 \delta \phi = \epsilon\partial_z\phi + \bepsilon \partial_{\bepsilon} \phi +\partial_z \epsilon +\partial_\bz \bepsilon +e^{-\phi} (\delta g_{\bz z}+\delta g_{z\bz}).
 \ee
 We insert this variation of $\phi$ into equation (\ref{EQ:polyakov}), and, assuming that integration by parts is legitimate,  
 reduce the terms involving $\epsilon$ to
 \bea
 &&-\frac{c}{12\pi} \int d^2x \,\partial_\bz \epsilon\left(\frac 12 (\partial_z \phi)^2-\partial^2_{zz} \phi\right)\nonumber\\
 &=&- \frac{c}{12 \pi}  \int d^2x\,  e^{-\phi} \delta g_{\bz\bz} \left(\partial^2_{zz} \phi- \frac 12 (\partial_z \phi)^2\right).
 \label{EQ:vary_polyakov1}
 \eea
 On comparing  with
 \bea
 \delta S_{\rm Polyakov}[g]&=&-\frac 12 \int d^2x\, \sqrt{g} \delta g_{\mu\nu} T^{\mu\nu}\nonumber\\ 
& =& -  \frac 12 \int d^2x\,  \sqrt{g}  \delta g_{\bz\bz}T^{\bz\bz},
   \eea
   where $\sqrt{g}\,d^2x =e^\phi dxdy$, 
   we read off that
   \bea
   \frac{c}{6\pi} e^{-2\phi}   \left(\partial^2_{zz} \phi- \frac 12 (\partial_z \phi)^2\right)&=& T^{\bz\bz}\nonumber\\
   &=& g^{\bz z}g^{\bz z}T_{zz}\nonumber\\
   &=& 4e^{-2\phi} T_{zz}.
   \eea
   Thus
\be
   T_{zz}= \frac {c}{24 \pi}  \left(\partial^2_{zz} \phi- \frac 12 (\partial_z \phi)^2\right).
 \ee
   Similarly we find that
   \be
     T_{\bz\bz}= \frac {c}{24 \pi} \left(\partial^2_{\bz\bz} \phi- \frac 12 (\partial_\bz \phi)^2\right).
   \ee
   
   Next, examining the effects of $\delta g_{\bz z}+ \delta g_{z\bz}$, we have
   \be
   \delta S_{\rm Polyakov}[g]= -\frac{c}{12\pi }  \int d^2x e^{-\phi} (\delta g_{\bz z}+ \delta g_{z\bz})(-\partial^2_{z\bz}\phi).
   \ee
   From this we read off that 
   \bea
   T^{z\bz}\,=\,T^{\bz z}&=& -\frac{c}{6  \pi }e^{-2\phi} \partial^2_{z\bz}\phi
   \nonumber\\
   &=& -\frac {c}{24  \pi} e^{-2\phi} \partial^2 \phi, 
     \eea
     and
     \be
     T_{z\bz}= -\frac{c}{24\pi }\partial_{z\bz}\phi.
     \ee 
     We also  recover the well-known trace anomaly \cite{duff}
   \be
    {T^\mu}_\mu = g_{\bz z}T^{\bz z}+ g_{z\bz}T^{\bz z}= e^{\phi} T^{z\bz}= \frac {c}{24\pi } R. 
    \label{EQ:trace_anom}
   \ee
   This is a comforting consistency check, as    Polyakov derived (\ref{EQ:polyakov}) by working backwards from (\ref{EQ:trace_anom}). 
   
We can now verify  that $T_{\mu\nu}$ is covariantly conserved:
   \bea
\frac 12 e^\phi   ( \nabla^z T_{zz}+\nabla^\bz T_{\bz z})&=& \nabla_\bz T_{zz}+\nabla_zT_{\bz z}\nonumber\\
&=& \partial_\bz T_{zz}+\partial_zT_{\bz z}-{\Gamma^z}_{zz}T_{\bz z}\nonumber\\
&=&   \partial_\bz T_{zz}+\partial_zT_{\bz z}-\partial_z \phi T_{\bz z}\nonumber\\
&=&0.
\label{EQ:conservedT}
\eea
This is evidence that  $S_{\rm Polyakov}[g]$ is indeed co-ordinate independent. There is a problem however: if     $S_{\rm Polyakov}$ is co-ordinate independent then  its functional derivative $T_{\mu\nu}$ must transform as a tensor.  When we make a holomorphic change of variables $z=z(\zeta)$, $\bz=\bz(\bzeta)$, however,   we have 
$
ds^2 = e^\chi d\zeta d\bzeta =e^{\phi}dz d\bz
$
and so
\be
\phi= \chi+\ln\left( \frac{\partial \zeta}{\partial z}\right) +\ln\left(\frac{\partial  \bzeta}{\partial \bz}\right).
\ee
Consequently  
\bea
T_{zz}&=&
\frac{c}{24\pi} \left(\partial^2_{zz} \phi- \frac 12 (\partial_z\phi)^2\right)\nonumber\\
&=&\frac{c}{24\pi} \left( \frac{\partial \zeta}{\partial z}\right)^2\left(\partial^2_{\zeta\zeta} \chi- \frac 12 (\partial_\zeta \chi)^2\right)+ \frac{c}{24\pi} \left(\frac{\zeta'''}{\zeta'} -\frac 32 \left(\frac{\zeta''}{\zeta'}\right)^2\right)\nonumber\\
&=&\left( \frac{\partial \zeta}{\partial z}\right)^2 T_{\zeta\zeta} + \frac{c}{24\pi} \{\zeta, z\},
\eea
where $T_{\zeta\zeta} $ is the energy-momentum tensor component evaluated in the $\zeta$, $\bzeta$ co-ordinates  and $\{\zeta, z\}$ is the Schwarzian derivative in whose definition the primes denote differentiation with respect to $z$.   Our $T_{\mu\nu}$ does {\it not\/} transform as a tensor therefore. The paradox is resolved by looking back at the first line in equation (\ref{EQ:vary_polyakov1}).  We see that if we are allowed to integrate by parts  we can take the $\partial_\bz$ derivative off of $\epsilon$ and onto $T_{zz}$. Thus any holomorphic addition to $T_{zz}$ is invisible to the variation $\delta g_{\bz\bz}$. Another way of saying this is that there can  be metric  variations $\delta g_{\bz\bz}$ that cannot be written  in the form  the form $\delta g_{\bz\bz}= -\partial_\bz \epsilon= -2\nabla_\bz \epsilon_\bz $. (The displacements $\epsilon$ and $\bepsilon$ should really be written as $\epsilon^z$ and $\epsilon^{\bz}$ as they are the components of a contravariant vector.)  The  solvability of (\ref{EQ:finding_epsilon}) depends on the global topology or on boundary conditions. On a torus, for example, metric variations due to change in the modular parameter $\tau$  are not expressible in this way.  On  a closed manifold of genus $g\ge 2$, there will be $3(g-1)$  linearly independent unobtainable metric variations.  

The addition of a purely holomorphic term is indeed required.  The full   {\it operator\/}  energy momentum tensor is 
\bea
\hat T_{zz}&=& \hat T(z)+      \frac {c}{24 \pi}  \left(\partial^2_{zz} \phi- \frac 12 (\partial_z \phi)^2\right),\nonumber\\
\hat T_{\bz\bz}&=& \hat {\bar T}(\bz)+      \frac {c}{24 \pi}  \left(\partial^2_{zz} \phi- \frac 12 (\partial_z \phi)^2\right),\nonumber\\
\hat T_{\bz z}&=& -\frac {c}{24  \pi} \partial^2_{z\bz} \phi,
\label{EQ:fullT}
\eea 
where, for a free $c=1$ boson field  $\varphi(z,\bz)= \varphi(z)+\varphi(\bz)$  for example, 
\bea
\hat T(z) &=&: \partial_z \varphi(z) \partial_z\varphi(z) :\nonumber\\
&=& \lim_{\delta  \to 0}\left( \partial_z \varphi(z+\delta/2) \partial_z \varphi(z-\delta/2)+ \frac{1}{4\pi \delta^2}\right). 
\label{EQ:Tz}
\eea 
(Note that conformal field theory  papers often  define $\hat T(z)$ to be $-2\pi$ times (\ref{EQ:Tz}) so as to simplify the operator product expansion.)
The operator $\hat T(z)$  has been constructed to be explicitly holomorphic,  but at a price of  tying its   definition  to the $z$, $\bz$ coordinate system --- both in  the mode normal ordering expression in  the first line  and by the explicit counterterm in the second. It is not surprising, therefore, that under  a holomorphic change of co-ordinates  the operator $\hat T(z)$ does not transform as a tensor. It  is well-known that instead
\be
\hat T(z) = \left( \frac{\partial \zeta}{\partial z}\right)^2 \hat T(\zeta)- \frac{c}{24\pi} \{\zeta,z\}.
\ee 
We see that  the  inhomogeneous Schwarzian derivative terms  cancel in the transformation of the   energy momentum tensor $\hat T_{\mu\nu}$ defined in  (\ref{EQ:fullT}). Thus $\hat T_{\mu\nu}$   transforms as a tensor and is still covariantly conserved.  It is notable that both the covariant conservation and the trace anomaly in $\hat T_{\mu\nu}$ are accounted for by the $c$-number terms. These properties   are therefore independent of the quantum state in which the expectation  is taken.  This quantum state only  influences the holomorphic part of  $\vev{\hat T_{zz}}$ and an antiholomorphic part of  $\vev{\hat T_{\bz\bz}}$.

In a {\it chiral\/} theory  we might constrain  both $\hat T_{\bz\bz}$ and $ \hat T_{\bz z}$ to be zero, while keeping the covariant form of $\hat T_{zz}$ defined in the first line of (\ref{EQ:fullT}).  The term $\nabla^\bz \hat T_{\bz z}$ needed for the continued mathematical validity of  (\ref{EQ:conservedT})  would then  be interpreted as
\bea
\nabla^\bz T_{\bz z} &\to & -\frac{c}{12\pi}\partial_z e^{-\phi}\partial_{z\bz}^2\phi\nonumber
\\
&=& \frac{c}{48\pi} \partial_z R,
\eea
so that  conservation law (\ref{EQ:conservedT})   is reinterpreted as the  anomaly equation appearing in \cite{cappelli-huerta-zemba}
\be
 \nabla^z \hat T_{zz} =- \frac{c}{48\pi}\partial_z R.
 \label{EQ:cappelli1}
 \ee
 By adding in an identically zero term we can write this as
 \be
 \nabla^z \hat T_{zz}+\nabla^\bz \hat T_{\bz z} =- \frac{c}{48\pi}\partial_z R,
 \label{EQ:cappelli2}
 \ee
 which at  first glance looks  like a covariant tensor equation. It is is not, however,  because  replacing the free index $z$ with $\bz$ leads to 
 \be
 \nabla^z \hat T_{z\bz}+ \nabla^\bz T_{\bz\bz}\stackrel{\rm ?}{=}- \frac{c}{48\pi}\partial_\bz R
 \ee
 on which the left hand side is identically zero, but the right need not be. Thus (\ref{EQ:cappelli2}) is {\it not\/} the covariant anomaly.
 
 A more symmetric treatment \cite{banerjee-kulkarni} 
 divides the trace anomaly between the left and right chiral sectors  and constrains  one of them to zero.  Then   $\hat T_{\bz\bz}$ remains  zero, but 
 \be
 \hat T_{z\bz}\to   -\frac{c}{48\pi} \partial^2_{z\bz}\phi,
 \ee
 so that 
 \be
 \hat {T^\mu}_\mu= \frac{c}{48\pi} R.
 \ee
 This physical reinterpretation  makes the (still {\it  mathematically\/} valid) equation   (\ref{EQ:conservedT}) read 
 \bea
  \nabla^z \hat T_{zz}+\nabla^\bz \hat T_{\bz z}=- \frac{c}{96\pi}\partial_z R, \\
  \nabla^z\hat T_{z\bz } +\nabla^\bz \hat T_{\bz \bz}= +\frac{c}{96\pi}\partial_\bz R,
 \eea
 where the second term on the left hand side of  the second equation is  constrained to be  zero.

 In our  $z$, $\bz$ co-ordinates system we have  $\sqrt{g} = \sqrt{-g_{\bz z}g_{z\bz}}=- i e^\phi/2$, and  we  can write these last two equations  in a covariant manner as   
\bea
  \nabla^z \hat T_{zz}+\nabla^\bz \hat T_{\bz z}=i \frac{c}{96\pi}\sqrt{g}\epsilon_{z\bz}  \partial^\bz R, \\
  \nabla^z\hat T_{z\bz } +\nabla^\bz \hat T_{\bz \bz}= i\frac{c}{96\pi}\sqrt{g}\epsilon_{\bz z} \partial^z R.
 \eea
 In general euclidean co-ordinates we therefore have \cite{fulling} 
 \be
 \nabla^\mu \hat T_{\mu\nu} = i\frac{c}{96 \pi} \sqrt{g} {\epsilon_{\nu \sigma}} \partial^\sigma R.
 \label{EQ:2dg_euclidanomaly}
 \ee
 The  factor ``$i$''  appears in (\ref{EQ:2dg_anomaly}) because it is only the {\it imaginary\/} part of the Euclidean effective action that can be anomalous \cite{alvarez-gaume-witten,alvarez-gaume-ginsparg}.  It is absent when we write the equation in Minkowksi signature space-time, where
 \be
 \nabla_\mu \hat T^{\mu\nu} = \frac{c}{96 \pi} \frac{1}{\sqrt{g} }{\epsilon^{\nu \sigma}} \partial_\sigma R.
 \label{EQ:2dg_anomaly}
 \ee
 
 Note that (\ref{EQ:2dg_anomaly}) can be rewritten as 
 $
 \nabla_\mu \widetilde T^{\mu\nu}=0
 $
 where 
 \be
 \widetilde T^{\mu\nu}= \hat T^{\mu\nu}-  \frac{c}{96 \pi} \frac{1}{\sqrt{g}} {\epsilon^{\nu \sigma}} R.
 \label{EQ:lorentz_current}
  \ee
  The new tensor $ \widetilde T^{\mu\nu}$ is conserved, but not symmetric. We have therefore exchanged a reparametrization anomaly for a Lorentz anomaly.

We now show that the    manifestly covariant  anomaly  (\ref{EQ:2dg_anomaly}) is  that expected from    the anomaly inflow.
 
 \section{Gravitational Chern-Simons terms}
 \label{SEC:gravCS}
 
 In this section we will use both the co-ordinate and frame-field (vielbein)  description of geometric quantities.  Thus   $e^\mu_a $ are the components of the  frame field  $\e_a= e^\mu_a\partial_\mu$ and $e^{\star b}_\mu$ the components of the co-frame $\e^{*a}= e^{*a}_\mu dx^\mu$, with  $\delta^a_b = e^{*a}_\mu e^\mu_b$.
The frame metric 
\be
\eta_{ab}= g_{\mu\nu} e^\mu_ae^\nu_b
\ee
is ${\rm diag}(1,1,1)$ and  ${\rm diag}(1,-1,-1)$  in Euclidean and  Minkowski space,  respectively.

 A gravitational $(2+1)$ dimensional Chern-Simons functional   can be written either in terms of the Christoffel-symbol   form ${\Gamma^\mu}_\nu= {\Gamma^\mu}_{\nu\sigma} dx^\sigma$as 
 \be
 C[\Gamma]= \frac{c}{96 \pi} \int_M \tr\{\Gamma d\Gamma+\frac 23 \Gamma^3\}.
 \ee
 or in terms of the spin connection ${\omega^a}_b={\omega^a}_{b\mu}dx^\mu$ as
 \be
 C[\omega]=  \frac{c}{96 \pi} \int_M \tr\{\omega d\omega+\frac 23 \omega^3\}.
 \ee
 The integrands in these two functionals have the same exterior derivative 
 \be
 d\,\tr\{\Gamma d\Gamma+\frac 23 \Gamma^3\}= d\,\tr\{\omega d\omega+\frac 23 \omega^3\}=\tr\{R^2\},
 \ee
and so they coincide when $M=\partial N$ is a boundary, but they are no longer  equal when $M$ itself has a boundary.
 Their normalization  is  related to the  index 
 \bea
 {\rm Index}(D_{\rm Dirac})&=&{\rm DimKer}(D_{\rm Dirac})- {\rm DimKer}(D^\dagger_{\rm Dirac})\nonumber\\
 &=& \frac{1}{192 \pi^2} \int_N \tr\{R^2\}
 \eea
 of the four dimensional Dirac operator. The Dirac index   an {\it even\/} integer for any   four-dimensional manifold possessing a spin structure.

 The spin connection is related to   the Christoffel  form by a ${\rm GL}(3)$ gauge transformation
 \be
{\omega^i}_{j\mu}= e^{*i}_\nu {\Gamma^\nu}_{\lambda\mu}e^\lambda_j  +e^{*i}_\nu \partial_\mu e^\nu_j, 
\ee
 and so
 \be
 C[\omega]=C[\Gamma] - \frac{c}{96\pi}  \int_{\partial M}\tr \{(de e^*) \Gamma\}- \frac{c}{288\pi} \int_M \tr \{(e^*de)^3\}.
 \ee
 Here the matrix-valued one-forms  $de e^*$ and $e^*de$ are defined by ${(de e^*)^\mu}_\nu \equiv (\partial_\sigma e_a^\mu) e^{*a}_\nu dx^\sigma$ and ${(e^*de)^a}_b\equiv  e^{*a}_\mu \partial_\sigma e^\mu_b d x^\sigma$.

  
The functional $C[\Gamma]$ is invariant under reparametrization $x^\mu\to X^\mu(x) $ up to boundary terms. To obtain   an energy-momentum conserving theory it has to be attached to a suitable boundary theory with compensating transformation properties. We do not have to write down the corresponding Wess-Zumino  action $W(\Gamma,X)$  to  know the boundary theory anomaly. All  we need to do is calculate the out-flowing  bulk energy-momentum  flux  by computing the response of 
$C[\Gamma]$ to a change in the metric.  
  
 The variation of the Chern-Simons functional due to a change in $\Gamma$ is 
 \be
 \delta C[\Gamma] = \frac{c}{48 \pi}  \int_M \tr\{\delta \Gamma R\}+\frac{c}{96 \pi}\int_{\partial M} \tr\{\delta \Gamma d\Gamma\}.
\ee
To compute the contribution to the energy-momentum tensor we also need 
\be
\delta {\Gamma^\mu}_{\nu\sigma}= \frac 12 g^{\mu\lambda}(\nabla_\nu \delta g_{\lambda\sigma}+ \nabla_\sigma \delta g_{\sigma \lambda}-\nabla_\lambda \delta g_{\nu\sigma}).
\ee
Then, making use of properties of the Riemann tensor that are unique to three dimensions (See \cite{krauss-larsen,perez}  for more details), we find  
\be
\delta C[\Gamma]= \frac{c}{48 \pi}  \int_M d^3 x \sqrt{g}C^{\mu\nu}\delta g_{\mu\nu} +\hbox{boundary terms},
\ee
where
\be
C^{\mu\nu} = - \frac{1}{2\sqrt{g}} \left(\epsilon^{\rho\sigma \mu}\nabla_\rho R^\nu_\sigma + \epsilon^{\rho\sigma \nu}\nabla_\rho R^\mu_\sigma\right)
\label{EQ:cotton}
\ee
is the {\it Cotton tensor\/}.  We read-off  the bulk energy-momentum tensor to be
\be
T^{\mu\nu}= -\frac{c}{24\pi}  C^{\mu\nu}
\ee
 In deriving this result we have had to integrate by parts a second time so as to  remove the derivatives from the metric variations. Consequently  the boundary terms are more complicated than the usual ones arising from the variation of gauge field Chern-Simons functionals. 
 We are, however,  confident that these boundary terms   provide the same  conversion of the consistent anomaly of the boundary theory into the covariant anomaly that we   saw with the gauge anomalies. 

We  restrict ourselves to product metrics of the form
\be
ds^2 = (dx^2)^2 +g_{a b}(x^0,x^1)dx^a dx^b,\quad a,b=0,1
\ee
with  boundary  being at  $x^2=0$.
The Ricci tensor apearing in (\ref{EQ:cotton})  then coincides with   the Ricci tensor of the two-dimensional boundary, and  can be written as 
\be
R^a_b= \frac 12 \delta^a_b R(x^0,x^1), \quad a,b=0,1
\ee
The flux of the $a=0, 1$ energy-momentum components into the boundary  becomes
\be
T^{2a} =  \frac{c}{96 \pi} \frac{1}{\sqrt{g}} \epsilon^{\rho a 2}\partial_\rho   R.
 \ee
 The energy momentum inflow into the boundary therefore precisely accounts for the the gravitational anomaly  (\ref{EQ:2dg_anomaly}). The ``suitable boundary theory'' is thus exactly the chiral theory whose anomaly we obtained in the previous section. 
 
 In contrast to $C[\Gamma]$, the  Chern-Simons functional $C[\omega]$  is  reparametrization invariant, but it  fails by boundary terms to be invariant under rotations (or Lorentz transformations) of the frame field:
 \bea
 \e_a &\to& \e_a^O= \e_b \,{O^b}_a,\nonumber\\
 {\omega^a}_b &\to&{(\omega^O)^{a}}_b= {(O^{-1})^a}_c {\omega^c}_d {O^d}_b + {(O^{-1})^a}_c d{O^c}_b.
 \eea
 To obtain the energy momentum flow associated with $C[\omega]$ we should  remember that  $\omega$ is linked to the  metric  through the torsion-free condition 
 \be
 d\e^{*a} +{{\omega}^a}_b \wedge \e^{*b}=0.
 \ee
 and through $g_{\mu\nu}= \eta_{ab}e^{*a}_\mu e^{*b}_\nu$.
 We therefore define a tensor
 $T_{bc}$ and its contravariant version $T^{da} = \eta^{db}\eta^{ac}T_{bc}$ by varying the vielbein:
\bea
\delta S_{\rm eff}&=& \int d^nx \sqrt{g}\left( \frac{\delta S}{\delta e^\mu_a}\right) \delta e^\mu_a\nonumber\\
&\equiv & \int d^nx \sqrt{g}\left(T_{bc} \eta^{ca} e^{*b}_\mu\right) \delta e^\mu_a.\nonumber\\
&=&  \int d^nx  \sqrt{g}\,T^{da}\delta e_{da}.
\eea
The last line introduces the useful quantity.
$\delta e_{da}= \eta_{db } e^{*b}_\mu \delta e^\mu_a$.
As defined, there is no immediate reason for $T_{bc}$ to be symmetric. 
However when the  functional $S$ is  invariant under  an infinitesimal  local rotation $\delta e^\mu_a= e^\mu_b {\theta^{b}}_a$,
we have
\bea
0&=&\delta S_{\rm eff}\nonumber\\
&=&   \int d^nx\sqrt{g} \,T_{bc}\,\eta^{ca} e^{*b}_\mu e^\mu_d {\theta^d}_a\nonumber\\ 
&=&  \int d^nx\sqrt{g} \,T_{bc}\,\eta^{ca}  {\theta^b}_a\nonumber\\
&=&  \int d^nx\sqrt{g}\, T^{da}\, \theta_{da}.\nonumber
\eea
Since $\theta_{da}$ is an arbitrary skew symmetric matrix, we see that $T^{da}=T^{ad}$.
Accepting this symmetry, we can now set
\bea
\delta{S}_{\rm eff}&=& \frac 12 \int d^nx\sqrt{g}\, T_{bc} \left(\eta^{ca} e^{*b}_\mu \delta e^\mu_a+\eta^{ba} e^{*c}_\mu \delta e^\mu_a \right)\nonumber\\
&=&\frac 12  \int d^nx\sqrt{g}\, T_{\alpha\beta}\left(e^\beta_c \eta^{ca}  \delta e^\alpha_a+e^\alpha_b \eta^{bc} \delta e^\beta_c \right)\nonumber\\
&=&\frac 12  \int d^nx\sqrt{g} \,T_{\alpha\beta}\, \delta g^{\alpha\beta}.\nonumber
\eea
Here   $T_{\alpha\beta}= e^b_\alpha e^c_\beta T_{bc}$.   Thus, for rotation invariant actions, the vielbein variation leads to the same energy-momentum tensor as Hilbert's   metric variation.

Now we have
 \be
 \delta C[\omega] = \frac{c}{48 \pi}  \int_M \tr\{\delta \omega R\}+\frac{c}{96 \pi}\int_{\partial M} \tr\{\delta \omega d\omega\}.
\ee
and we can use 
\bea
(\delta \omega_{ij\mu}) e^\mu_k 
& =&-\frac 12\left\{(\nabla_j \delta e_{ik} -   \nabla_k  \delta e_{ij})\right.\nonumber\\
&&\quad+( \nabla_k \delta e_{ji}-   \nabla_i  \delta e_{jk})\nonumber\\
&&\quad \left.-( \nabla_i \delta e_{kj}-   \nabla_j \delta e_{ki})\right\}\nonumber
\eea
to compute $T_{ab}$.  We do not have to perform this rather tedious computation, however. We know that the variations of  $C[\Gamma]$ and $C[\omega]$ differ only by boundary terms. The {\it bulk\/}  energy-momentum tensors for the two actions must therefore coincide. The boundary variations will  differ though. Because $C[\omega]$ is reparametrization invariant, the Wess-Zumino term
\be
W[\omega,O]\stackrel{\rm def}{=} C[\omega^O]-C[\omega]
\ee
that together with $C[\omega]$ gives the rotation  and reparametrization invariant action $C[\omega^O]$,
must give rise to a {\it conserved\/}  boundary-theory energy-momentum tensor $T^{ab}_{\rm WZ}$. This tensor must  also be covariant under co-ordinate changes, but will not be  symmetric.  There is only one possibility --- the frame field version of (\ref{EQ:lorentz_current}):
\be
T^{ab}_{\rm WZ}= \widetilde T^{ab}= \hat T^{ab} - \frac{c}{96\pi} \frac{1}{\sqrt{g}}\epsilon^{ab} R.
\ee
The  contribution $X^{ab}$ that comes from the boundary part of the variation of $C[\omega]$  will then repair the asymmetry. This contribution is easily computed, and is  
\be
X^{ab}= \frac{c}{96\pi}\frac{1}{\sqrt{g}} \epsilon^{ab} R.
\ee
The net effect is that we get the same boundary-theory energy-momentum tensor $\hat T^{\mu\nu}= T^{\mu\nu}_{\rm WZ}+X^{\mu\nu}$, and the same anomaly equation,  independent of whether  we write the gravitational Chern-Simons function in terms of $\Gamma$ or in terms of  $\omega$.   The only  difference between the two formulations lies in  the manner in which the boundary energy-momentum  is apportioned  between the  bulk Chern-Simons contribution  $X^{\mu\nu}$  and the boundary Wess-Zumino part $T^{\mu\nu}_{\rm WZ}$.

\section{Conclusions}
\label{SEC:conclusions}

We have seen that it is most likely that the thermal Hall currents on the surface of topological insulators are confined to one dimensional domain walls, and cannot flow in the two-dimensional surface.  To confirm this idea we computed the energy-momentum flux associated with a gravitational Chern-Simons term in the boundary effective action. We found  that the energy-momentum flux is proportional to gradients of the Ricci curvature, and therefore needs tidal forces to be non-zero.  We   related this flux to the gravitational anomaly experienced  by modes localized  on one-dimensional domain walls within the surface, and showed  that this anomaly takes the same covariant form independently of whether we write the gravitational Chern-Simons functional in terms of the Christoffel symbol $\Gamma$ or the spin connection $\omega$.

\section{acknowledgements}

This work was supported by   the National Science Foundation  under grant  DMR 09-03291. This work was begun during the TopoMat-11 program  at   KITP, Santa Barbara,  and was there  supported in part by the National Science Foundation under Grant No. NSF PHY05-51164.
I would like to thank Andreas Ludwig, Shinsei Ryu, and Taylor Hughes for  discussions, and also Joel Moore and Shoucheng Zhang for comments and feedback.

\end{document}